\documentclass[amsmath,floatfix,twocolumn,superscriptaddress]{revtex4-1}
\usepackage{amssymb,amsmath}
\usepackage{graphicx,array}
\usepackage{dcolumn}
\usepackage{psfrag}
\usepackage{bm}
\usepackage{color}

\begin{document}
\title{Is the normal to superionic transformation occurring in type-II fast-ion conductors a 
       real thermodynamic phase transition?}

\author{Claudio Cazorla}
\thanks{Corresponding Author}
\affiliation{School of Materials Science and Engineering, UNSW Sydney, NSW 2052, Australia}

\author{Daniel Errandonea}
\affiliation{Departament de F\'isica Aplicada, Institut de Ci\`encia de Materials, MALTA Consolider Team, Universitat de Val\`encia, 
             Edifici d'Investigaci\'o, Burjassot 46100, Spain}

\maketitle

Fedorov has written a Comment on our recent paper \cite{cazorla18} in which he suggests that the transformation
from a normal ($\alpha$) to a superionic ($\beta$) state occurring in type-II fast-ion conductors (e.g., CaF$_{2}$ and 
Li-based superionic materials \cite{hull04}) cannot be regarded as a phase transition in a rigorous thermodynamic 
sense. Fedorov's arguments for such a criticism can be summarized as follows: (1)~the available heat capacity data for the 
$\alpha \to \beta$ transformation do not reproduce the expected behaviour for a thermodynamic phase transition, (2)~there 
is not evidence for molar volume discontinuity during the superionic $\alpha \to \beta$ transformation, and (3)~the 
$\alpha \to \beta$ transformation is symmetry preserving. Besides these three tentative arguments, Fedorov's presents a 
fourth one based on the binary phase diagram of CaF$_{2}$--LaF$_{3}$ solid solutions, on which we will comment by the end 
of this Reply. 

In what follows, we demonstrate either by reproducing experimental data published by other authors or by explaining some 
simple solid-state arguments that the statements put forward by Fedorov are incorrect and do not correspond to reality. 
In fact, based on mounting experimental evidence, the normal to superionic phase transformation occurring in type-II 
fast-ion conductors can be rigorously considered, and should be referred to, as a phase transition. Notwithstanding,
it remains less obvious to establish with generality which is the order of the $\alpha \to \beta$ phase transition. 

\begin{figure}
\centerline{
\includegraphics[width=0.90\linewidth]{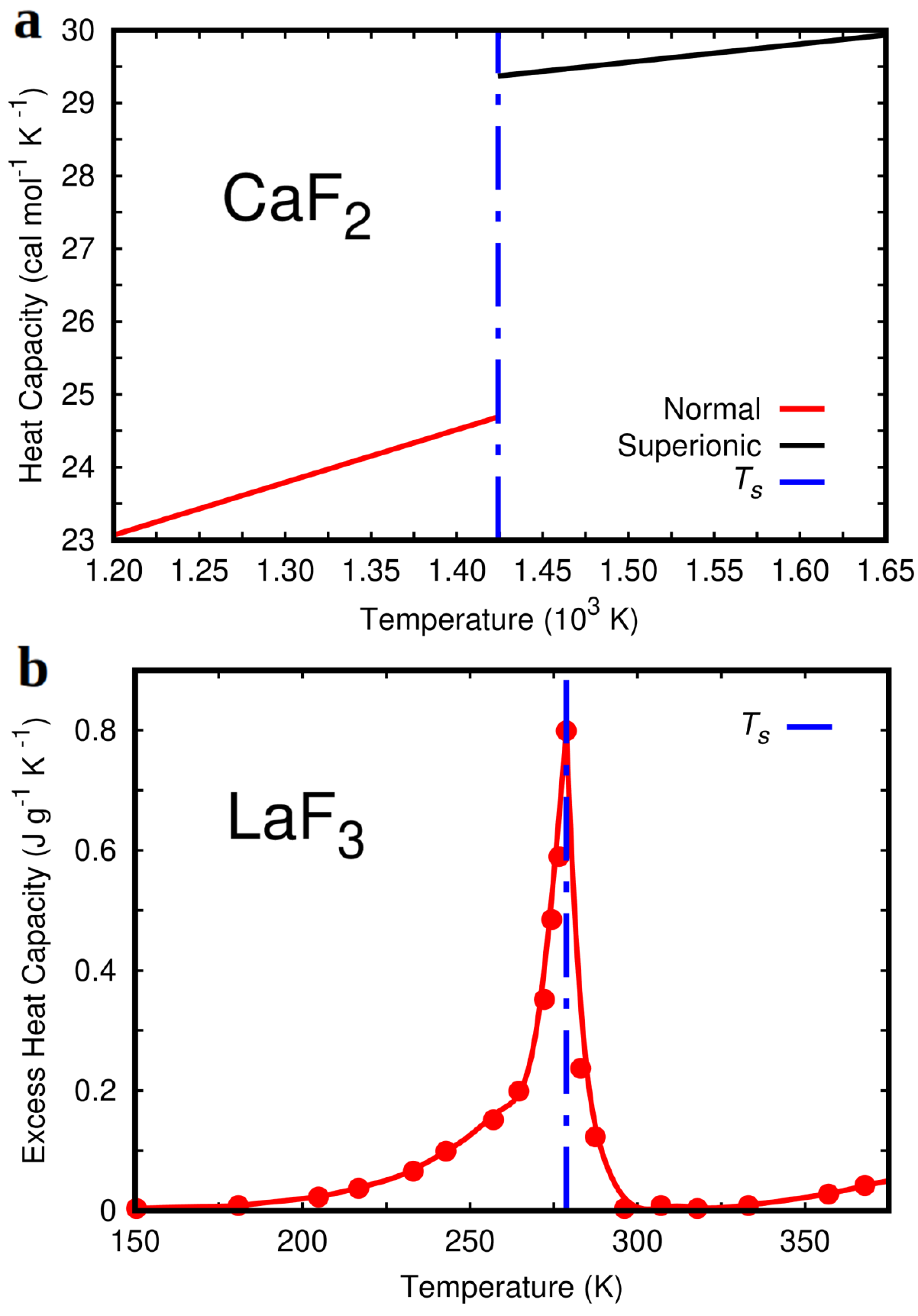}}
\caption{Experimental heat capacity of different type-II superionic materials. {\bf a}~Experimental data
        for CaF$_{2}$ reproduced from work \cite{naylor45}. The $C_{P}$ curve indicates that the $\alpha 
	\to \beta$ phase transition is of second-order type. {\bf b}~Experimental data for LaF$_{3}$ 
	reproduced from work \cite{krivorotov10}. The $C_{P}$ curve indicates that the $\alpha \to \beta$ 
	phase transition is of first-order type.}
\label{fig1}
\end{figure}

Before presenting our evidence, and to avoid possible misunderstandings, let us briefly review the thermodynamic 
behaviour of a pure substance that undergoes a first-order or a second-order phase transition. It is well known from 
thermodynamics that the heat capacity of a system at constant pressure, $C_{P}$, can be expressed as a function of the 
second derivative of its Gibbs free energy, $G$, as:
\begin{equation}
C_{P} \equiv T \left( \frac{\partial S}{\partial T} \right)_{P} = - T \left( \frac{\partial^{2} G}{\partial T^{2}} \right)_{P}~,
\label{eq:cp}
\end{equation}
where $P$, $T$ and $S$ represent the pressure, temperature and entropy of the system, respectively, and the relation $S = 
-\left( \frac{\partial G}{\partial T} \right)_{P}$ has been used. By definition, first-order phase transitions involve 
discontinuities in the first derivative of $G$ at the transition point (e.g., in the system entropy). Consequently,  
$C_{P}$ exhibits a singularity at the phase-transition point, which in practice is measured in the form of a sudden 
increase or ``bump'' (recall that ``infinity'' is an abstraction that in practice can never be measured). Meanwhile, 
and also by definition, second-order phase transitions present discontinuities in the second derivative of $G$ at the transition 
point. Consequently, $C_{P}$ presents a discontinuity at the phase-transition point (Eq.\ref{eq:cp}). Likewise, at a 
fixed pressure the presence of a sudden increase or ``bump'' on the measured thermal expansion of a compound, defined as: 
\begin{equation}
\kappa_{P} \equiv \frac{1}{V} \left( \frac{\partial V}{\partial T} \right)_{P}~,
\label{eq:thermal}
\end{equation}
indicates that the volume of the system, $V$, changes discontinuously at the corresponding temperature. 

\begin{figure}
\centerline{
\includegraphics[width=0.90\linewidth]{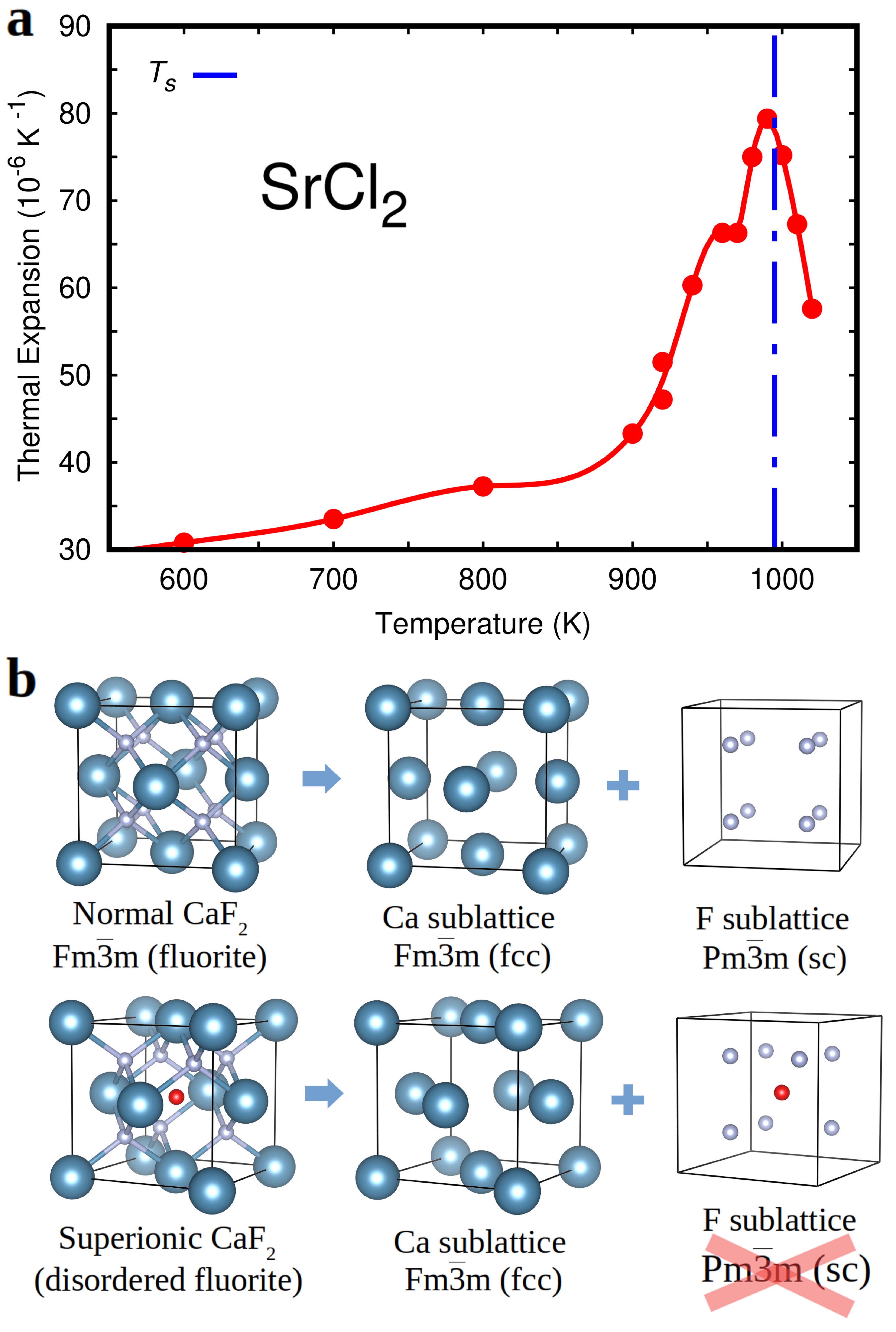}}
\caption{{\bf a}~Experimental thermal expansion of fluorite SrCl$_{2}$ reproduced from work
        \cite{moore85}. The $\kappa_{P}$ function indicates that during the $\alpha \to \beta$
        phase transition the volume of the system changes discontinuously. {\bf b}~Ball-stick representation of
        the normal and superionic CaF$_{2}$ phases. The crystal structure of the $\alpha$ and $\beta$
        phases are different due to the presence of F interstitials in the disordered fluorite phase. Ca and
        F ions in regular fluorite positions are represented with blue and violet spheres; F interstitial ions in
        off-symmetry fluorite positions are represented with red spheres.}
\label{fig2}
\end{figure}

Figure~\ref{fig1}a shows the experimental heat capacity of CaF$_{2}$ at high temperatures and zero pressure 
\cite{naylor45}. Fluorite CaF$_{2}$ is an archetypal type-II superionic material that undergoes a normal to 
superionic transformation at temperature $T_{s} \approx 1400$~K \cite{hull04,naylor45,cazorla13,voronin01,cazorla16,cazorla14,cazorla17}. 
As per our thermodynamic explanations above, the shape of the $C_{P}$ curve measured for CaF$_{2}$ corresponds to a 
second-order phase transition because it displays a discontinuity at $T_{s}$. Moreover, Fig.\ref{fig1}b shows 
the experimental $C_{P}$ of LaF$_{3}$ \cite{krivorotov10}, another well-known type-II superionic material 
that despite not presenting the fluorite structure appears mentioned in Fedorov's Comment. In this case, the 
sudden increase of the heat capacity at the superionic transition temperature indicates that the normal 
to superionic phase transition is of first-order type. 

In the scientific literature, one can certainly find abundant examples of fluorite type-II superionic materials 
that display clear $C_{P}$ or $S$ discontinuities at their corresponding superionic transition temperatures. For instance, 
the entropies of the superionic transition reported for SrCl$_{2}$, PbF$_{2}$, BaF$_{2}$ and BaCl$_{2}$ in the 
experimental work \cite{schroter80} strongly suggest that their $\alpha \to \beta$ phase transitions are 
of first-order type (e.g., see Figs.3--4 in \cite{schroter80}). A very similar experimental behaviour has 
been also reported for the anti-fluorite fast-ion conductor Cu$_{2-x}$Se \cite{kang16}. Therefore, Fedorov's 
statement (1) appears to be totally flawed and unrealistic. It is noted, however, that it does not seem possible 
to generalize the order of the superionic $\alpha \to \beta$ phase transition. 

Figure~\ref{fig2}a shows the experimental thermal expansion of fluorite SrCl$_{2}$ at high temperatures
and zero pressure \cite{moore85}. The shape of the $\kappa_{P}$ curve indicates the presence of a volume 
discontinuity at the corresponding superionic transition. Similarly, $\kappa_{P}$ experimental data 
supporting the existence of a phase transition in other fluorite compounds like CaF$_{2}$, PbF$_{2}$, 
SrF$_{2}$ and BaF$_{2}$ can be also found in work \cite{roberts86}. Furthermore, in this latter reference it is 
explicitly shown that the observed ``bumps'' of the heat capacity and thermal expansion occur both at 
$T_{s}$ \cite{roberts86}, which is another clear phase-transition manifestation. Therefore, it can be 
concluded that Fedorov's statement (2) is also incorrect.  

It is very well known from neutron scattering experiments \cite{hutchings84} and atomistic computer simulations 
\cite{cazorla17,gillan86,gillan90,dent04} that ionic transport in fluorite type-II superionic conductors occurs via 
discrete anion jumps that involve occupancy of off-symmetry fluorite positions (Fig.\ref{fig2}b). In particular,  
the sublattice formed by the anions in the superionic $\beta$ phase is different from the simple cubic arrangement 
observed in the normal $\alpha$ phase (Fig.\ref{fig2}b). Therefore, in a strict sense the $\alpha \to \beta$ phase 
transition cannot be regarded as symmetry preserving because the space group of the disordered fluorite phase is 
not $Fm\overline{3}m$ (Fig.\ref{fig2}b). Consequently, Fedorov's statement (3) neither is correct.  

Finally, Fedorov presents an extravagant argument based on the binary phase diagram of CaF$_{2}$--LaF$_{3}$ solid solutions.  
LaF$_{3}$ exhibits a rhombohedral $P\overline{3}c1$ structure that differs greatly from the cubic $Fm\overline{3}m$ 
phase. The ionic diffusion mechanisms and miscibility properties of such non-fluorite CaF$_{2}$--LaF$_{3}$ solid solutions 
are far from the focus of our work \cite{cazorla18}, which is on pure fluorite superionic compounds. Therefore, due 
to its inadequacy, we simply ignore this argument. 

In conclusion, in this Reply we have demonstrated that contrarily to Fedorov's arguments the superionic $\alpha \to \beta$ 
transformation occurring in fluorite type-II fast-ion conductors can be considered as a phase transition in a rigorous
thermodynamic sense. As a matter of fact, researchers have judiciously referred to this transformation as ``superionic
phase transition'' for more than three decades \cite{hull04,naylor45,cazorla13,voronin01,cazorla16,cazorla14,cazorla17,krivorotov10,
schroter80,kang16,moore85,roberts86,hutchings84,gillan86,gillan90,dent04}, just as we did in our recent work \cite{cazorla18}.

\end{document}